\documentclass[12pt,twoside]{article}
\usepackage{wrapfig}
\usepackage{graphicx}
\usepackage{cmp2e}
\title[Reentrant transitions of a mixed-spin Ising model]
{Reentrant transitions of a mixed-spin Ising model on the diced lattice\thanks{
This work was financially supported under the grants VEGA 1/2009/05 and APVT 20-005204.}}
\author{M. Ja\v{s}\v{c}ur, J. Stre\v{c}ka}
\address{Department of Theoretical Physics and Astrophysics, \\
Faculty of Science, P. J. \v{S}af\'arik University, \\ 
Park Angelinum 9, 040 01 Ko\v{s}ice, Slovak Republic}
\begin{document}

\maketitle

\begin{abstract}

Magnetic behaviour of a mixed spin-1/2 and spin-1 Ising model on the diced lattice is studied 
by the use of an exact star-triangle mapping transformation. It is found that the uniaxial as 
well as biaxial single-ion anisotropy acting on the spin-1 sites may potentially cause a 
reentrant transition with two consecutive critical points. Contrary to this, the effect of 
next-nearest-neighbour interaction between the spin-1/2 sites possibly leads to a reentrant 
transition with three critical temperatures in addition to the one with two critical points only. 
The shape of the total magnetization versus temperature dependence is particularly investigated 
for the case of ferrimagnetically ordered system.

\pacs 75.10.Hk, 05.50.+q, 75.40
\end{abstract}

\section{Introduction}

Over the last six decades, much effort has been devoted to determine a criticality and other statistical properties of various lattice-statistical models, which would enable a deeper understanding of order-disorder phenomena in solids. The planar Ising model takes a prominent place in the equilibrium statistical physics from this point of view as it represents a rare example of exactly solvable lattice-statistical model since Onsager's pioneering work \cite{Ons44}. It is noteworthy that an exact tractability of two-dimensional Ising systems \cite{Bax82} is being of immense importance because of providing a convincing evidence for many controversial results predicted in the phase transition theory and moreover, the exactly soluble models  provide a good testing ground for approximative theories. It should be emphasized, however, that a precise treatment of planar Ising models is usually connected with the usage of sophisticated mathematical methods, which consequently lead to considerable difficulties when applying them to more complicated but simultaneously 
more realistic systems.   

The Ising systems consisting of mixed spins of different magnitudes, so-called {\it mixed-spin Ising models}, are being among the most interesting extensions of the standard spin-1/2 Ising model which are currently of great research interest. These models enjoy a considerable attention mainly due to much richer critical behaviour they display compared with their single-spin counterparts. Magnetic properties of the mixed-spin Ising models can essentially be modified, for instance, by a presence of single-ion anisotropy acting on the $S \geq 1$ spins. Indeed, this effect induced by a crystal field of ligands may potentially cause a tricritical phenomenon, a change of the character of phase transition from a second-order to a first-order one. Another aspect which started to attract an appreciable interest towards the mixed-spin Ising models can be related to a theoretical modeling of magnetic structures suitable for describing a ferrimagnetism of certain class of insulating materials. Accordingly, the ferrimagnetic mixed-spin Ising models are very interesting also from the experimental viewpoint, first of all in connection with many 
possible technological applications of ferrimagnets to practice (e.g. thermomagnetic recording).

Despite the intensive studies, there are only few examples of exactly solvable mixed-spin models, yet. 
Using the generalized form of decoration-iteration and star-triangle transformations, the mixed spin-1/2 
and spin-$S$ ($S \geq 1$) Ising models on the honeycomb, diced and decorated honeycomb lattices \cite{Fis59}, as well as on the decorated square, triangular and diced lattices \cite{Yam69} were exactly been treated 
long ago. Notice that these mapping transformations have been later on further generalized in order to 
account also for the single-ion anisotropy effect. The influence of uniaxial single-ion anisotropy on 
magnetic properties of the mixed-spin systems have been precisely investigated on the honeycomb lattice \cite{Gon85} and on some decorated planar lattices \cite{Jas98}. In addition, more general form of 
the star-triangle transformation regarding the biaxial single-ion anisotropy have only recently been 
employed to obtain exact results for the mixed-spin Ising model on the honeycomb lattice \cite{Str04}. 
To the best of our knowledge, these are the only mixed-spin systems with generally known exact solutions except the ones on the Bethe (Cayley tree) lattices, which can be studied within a discrete non-linear map \cite{Sil91} and/or an approach based on exact recursion equations \cite{Alb03}. Among the remarkable models for which a precise solution is restricted to a certain subspace of interaction parameters only, one 
should further mention the mixed-spin Ising models on the union jack (centered square) lattice 
that can be mapped onto the eight-vertex model \cite{Lip95}.  

The purpose of this work is to provide an exact formulation of the mixed spin-1/2 and spin-1 Ising model 
on the diced lattice and to establish accurate phase diagrams of this system in dependence on a strength 
of the uniaxial and biaxial single-ion anisotropies as well as on a strength of the next-nearest-neighbour interaction. Exact results of this system are obtained by applying the generalized star-triangle transformation constituting an exact correspondence with the effective spin-1/2 Ising model 
on the triangular lattice. Last, the effect of single-ion anisotropy terms and the next-nearest-neighbour interaction on the temperature dependence of the resultant magnetization will be clarified for the case 
of ferrimagnetic system. 

The organization of this  paper is as follows. In Section 2, a detailed formulation of the model 
system is presented and subsequently, exact expressions for the magnetization and critical 
temperatures are derived by establishing an exact correspondence with the effective spin-1/2 
Ising model on the triangular lattice. The most interesting numerical results are presented 
and particularly discussed in Section 3. Finally, some concluding remarks are given in Section 4.
\begin{figure}
\centerline{\includegraphics[width=0.93\textwidth]{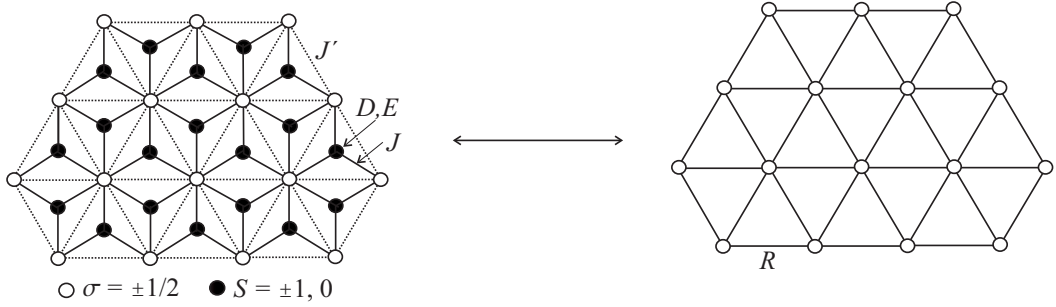}}
\caption{Diagrammatic representation of the diced lattice consisting of the mixed spin-1/2 (empty circles)
and spin-1 (solid circles) sites, respectively. The solid (broken) lines depict the nearest-neighbour (next-nearest-neighbour) interactions. By the use of star-triangle transformation, the mixed-spin diced 
lattice is mapped onto a simple spin-1/2 triangular lattice.}
\label{fig1}
\end{figure}

\section{Model and its exact solution}

Let us begin by considering the mixed spin-1/2 and spin-$1$ Ising model on the diced lattice schematically illustrated on the left-hand side of figure 1. The mixed-spin diced lattice consists of two interpenetrating sub-lattices $A$ and $B$ which are formed by the spin-1/2 (empty circles) and spin-1 (solid circles) atoms, respectively. The Ising Hamiltonian defined upon the underlying lattice reads:
\begin{eqnarray}
{\hat \mathcal H}_{diced} = - J  \sum_{(i,j) \subset \mathcal L}^{6N} \hat S_{i}^z \hat \sigma_{j}^z
     - J' \sum_{(k,l) \subset \mathcal K}^{3N} \hat \sigma_{k}^z \hat \sigma_{l}^z
     - D \sum_{i=1}^{2N} (\hat S_{i}^z)^2 - E \sum_{i=1}^{2N} [(\hat S_{i}^{x})^2 - (\hat S_{i}^{y})^2],     
\label{HD}
\end{eqnarray}
where $\hat \sigma_j^z$ and $\hat S_i^{\alpha}$ ($\alpha = x, y, z$) stand for the spatial components 
of spin-1/2 and spin-1 operators, respectively, $J$ denotes the exchange interaction between the
nearest-neighbouring spin pairs and $J'$ labels the interaction between the next-nearest-neighbouring
spin-1/2 atoms. Finally, the parameters $D$ and $E$ measure a strength of the uniaxial and biaxial single-ion anisotropies acting on the spin-1 sites and $N$ denotes the total number of the spin-1/2 sites. 

It can be clearly seen from figure 1 that the considered diced lattice belongs to loose-packed lattices, 
i.e. it can be decomposed into two interpenetrating sub-lattices $A$ and $B$ in such a way that all 
nearest neighbours of the spin-1 sites are being the spin-1/2 sites and vice versa. Consequently, 
a trace over spin degrees of freedom of all spin-1 atoms can be performed before summing over all 
available configurations of the spin-1/2 atoms. It is therefore very convenient to write the 
total Hamiltonian (\ref{HD}) as a sum of the commuting site Hamiltonians ${\hat \mathcal H}_{diced} = \sum_{i} {\hat \mathcal H}_{i}$, where each site Hamiltonian ${\hat \mathcal H}_{i}$ involves all interaction terms associated with one appropriate spin-1 site residing $i$th lattice point of the sub-lattice $B$:
\begin{eqnarray}
{\hat \mathcal H}_{i} = \! \! &-& \! \! \hat S_{i}^z J (\hat \sigma_{i1}^z + \hat \sigma_{i2}^z + 
                \hat \sigma_{i3}^z) - (\hat S_{i}^z)^2  D - [(\hat S_{i}^{x})^2 - (\hat S_{i}^{y})^2] E 
\nonumber \\ \! \! &-& \! \!  J' (\hat \sigma_{i1}^z \hat \sigma_{i2}^z + \hat \sigma_{i2}^z 
                                  \hat \sigma_{i3}^z + \hat \sigma_{i3}^z \hat \sigma_{i1}^z)/2
\label{HI}
\end{eqnarray}
The factor $\frac{1}{2}$ emerging in the last expression arises in order to avoid a double counting 
of each next-nearest-neighbour interaction  when summing over all spin-1 sites. On account of the commutability between the site Hamiltonians (\ref{HI}), the partition function of the mixed-spin 
diced lattice can be partially factorized to:
\begin{eqnarray}
{\mathcal Z}_{diced} = 
\mbox{Tr}_{\{\sigma \}} \prod_{i=1}^{2N} \mbox{Tr}_{S_i} \exp(- \beta {\hat \mathcal H}_{i}),     
\label{ZD}
\end{eqnarray}
where $\beta = 1/(k_{\mathrm B} T)$, $k_{\mathrm B}$ is being Boltzmann's constant, $T$ stands for the absolute temperature, the symbol $\mbox{Tr}_{S_i}$ means a trace over the spin degrees of freedom of 
$i$th spin-1 atom of the sub-lattice $B$ and the symbol $\mbox{Tr}_{\{\sigma \}}$ stands for a trace 
over all possible spin configurations of the spin-1/2 atoms on the sub-lattice $A$. 

After performing elementary calculations (see for instance reference \cite{Str04}), one readily finds an analytical expression for $\mbox{Tr}_{S_i} \exp(- \beta {\hat \mathcal H}_{i})$. Moreover, an explicit form of 
the relevant trace immediately implies a possibility of performing the familiar star-triangle mapping transformation \cite{Fis59}:
\begin{eqnarray}
\mbox{Tr}_{S_i} \exp(- \beta {\hat \mathcal H}_{i}) \! \! &=& \! \! \exp[\beta J'(\sigma_{i1}^z \sigma_{i2}^z + \sigma_{i2}^z \sigma_{i3}^z + \sigma_{i3}^z \sigma_{i1}^z)/2)] \nonumber \\ \Bigl \{ 1 \! \! &+& \! \! 2 \exp(\beta D) \cosh \Bigl[\beta \sqrt{J^2 (\sigma_{i1}^z + \sigma_{i2}^z + \sigma_{i3}^z)^2 + E^2} \Bigr] \Bigr \} \nonumber \\ \! \! &=& \! \! 
C \exp[\beta R (\sigma_{i1}^z \sigma_{i2}^z + \sigma_{i2}^z \sigma_{i3}^z + \sigma_{i3}^z \sigma_{i1}^z)/2]
\label{ST}
\end{eqnarray}
which establishes an exact correspondence between the mixed-spin Ising model on the diced lattice and 
its equivalent spin-1/2 Ising model on the triangular lattice provided that the transformation (\ref{ST}) 
is applied to all spin-1 sites (see figure 1). Notice that the unknown mapping parameters $C$ and $R$ are unambiguously given directly by the mapping transformation (\ref{ST}), which must hold for all available configurations of $\sigma_{i1}^z$, $\sigma_{i2}^z$, $\sigma_{i3}^z$ spins. According to this, one easily 
finds:
\begin{eqnarray}
C^4 = V_1 V_2^{3}, \qquad \quad \beta R = \beta J' + 2 \ln(V_1/V_2), 
\label{MP}
\end{eqnarray}
with the expressions $V_1$ and $V_2$ defined as:
\begin{eqnarray}
V_1 &=&  1 + 2 \exp(\beta D) \cosh \Bigl( \beta \sqrt{(3J/2)^2 + E^2} \Bigr),  \label{MP1}   \\
V_2 &=&  1 + 2 \exp(\beta D) \cosh \Bigl( \beta \sqrt{(J/2)^2 + E^2} \Bigr).  \label{MP2}
\end{eqnarray}

Now, let us substitute the mapping relation (\ref{ST}) into the formula (\ref{ZD}) gained 
for the partition function of the mixed-spin diced lattice. After straightforward rearrangement, 
one easily obtains an exact relationship between the partition function ${\mathcal Z}_{diced}$ 
of the mixed-spin Ising model on the diced lattice and respectively, the partition function 
${\mathcal Z}_{triang}$ of its corresponding spin-1/2 Ising model on the triangular lattice:
\begin{eqnarray}
{\mathcal Z}_{diced} (\beta, J, J', D, E) = C^{2N} {\mathcal Z}_{triang} (\beta, R).
\label{PF}
\end{eqnarray}
With regard to this, the mixed-spin diced lattice is mapped onto the spin-1/2 Ising model on the 
triangular lattice exactly solved several years ago \cite{Hou50}-\cite{Bax89}. 

The mapping relation (\ref{PF}) between the partition functions ${\mathcal Z}_{diced}$ and ${\mathcal Z}_{triang}$ represents a central result of our calculation, in fact, this relation can be utilized for establishing similar mapping relations also for other important quantities such as Gibbs free energy, internal energy, magnetization, correlation functions, specific heat, etc. When combining the formula (\ref{PF}) with 
commonly used mapping theorems \cite{Bar88}, one readily proves a validity of following relations:
\begin{eqnarray}
m_A &\equiv& \langle \hat \sigma_i^z \rangle_{diced} = 
\langle \hat \sigma_i^z \rangle_{triang} \equiv m_{triang},  \label{mA} \\
t_A &\equiv& \langle \hat \sigma_i^z \hat \sigma_j^z \hat \sigma_k^z \rangle_{diced} = 
     \langle \hat \sigma_i^z \hat \sigma_j^z \hat \sigma_k^z \rangle_{triang} \equiv t_{triang}, \label{tA} \\
m_B  &\equiv& \langle S_i^z \rangle_{diced} = 3 m_A [F(3) + F(1)]/2 + 2 t_A [F(3) - 3F(1)],                   
\label{mB}
\end{eqnarray}
where the symbols $\langle ... \rangle_{diced}$ and $\langle ... \rangle_{triang}$ denote standard canonical average performed over the Ising models on the mixed spin-1/2 and spin-1 diced lattice and on the spin-1/2 triangular lattice, respectively, and the function $F(x)$ stands for:
\begin{eqnarray}
F(x) =  \frac{Jx/2}{\sqrt{(Jx/2)^2 + E^2}} \frac{2 \sinh(\beta \sqrt{(Jx/2)^2 + E^2})}
                                                {\exp(- \beta D) + 2 \cosh(\beta \sqrt{(Jx/2)^2 + E^2})}. 
\label{F}
\end{eqnarray}
It should be pointed out that the set of equations (\ref{mA})-(\ref{mB}) enables straightforward calculation of both sub-lattice magnetization $m_A$ and $m_B$ of the mixed-spin system on the diced lattice. 
According to equation (\ref{mA}), the sub-lattice magnetization $m_A$ directly equals to the 
corresponding magnetization $m_{triang}$ of the triangular lattice, which is being unambiguously given by 
the mapping relation (\ref{MP}). Note that the exact result for magnetization of the spin-1/2 Ising model 
on the triangular lattice was firstly derived by Potts in his short note \cite{Pot52}:
\begin{eqnarray}
m_{triang} =  \frac{1}{2} \Bigl[ 1 - \frac{16 x^3}{(1 + 3x)(1 - x)^3}\Bigr]^{1/8}, 
\label{M}
\end{eqnarray}
where $x = \exp(- \beta R)$. On the other hand, an exact knowledge of the triplet order parameter $t_{triang}$ is being required in order to compute the sub-lattice magnetization $m_B$ in addition to the single-site magnetization $m_{triang}$. Exact results of Baxter and Choy \cite{Bax89} for the anisotropic triangular 
Ising model include, as specialization, the following exact solution for the triplet correlation of 
the isotropic triangular lattice:
\begin{eqnarray}
t_{triang} =  \frac{m_{triang}}{4}  
              \biggl[1 + 2 \frac{x^2 - 2x^{-2} + 1 - \sqrt{(x^2 + 3)(x - 1)}}{(x - x^{-1})^2} \biggr]. 
\label{T}
\end{eqnarray}
In view of the results (\ref{M}) and (\ref{T}), an exact solution of both sub-lattice magnetization 
of the mixed-spin system is thus formally completed.  

Finally, let us derive an analytical condition that determines a criticality of the mixed-spin diced 
lattice. It can easily be understood that both sub-lattice magnetization $m_A$ and $m_B$ vanish if
the magnetization $m_{triang}$ of the corresponding triangular lattice tends to zero. It is therefore 
sufficient to substitute an exact critical point of the isotropic triangular lattice \cite{Hou50} 
into the mapping relation (\ref{MP}) in order to obtain an accurate condition for the criticality 
of the mixed-spin Ising model:
\begin{eqnarray}
\beta_c J' + 2 \ln \Biggl \{ \frac{1 + 2 \exp(\beta_c D) \cosh \Bigl( \beta_c \sqrt{(3J/2)^2 + E^2} \Bigr)}
                                {1 + 2 \exp(\beta_c D) \cosh \Bigl( \beta_c \sqrt{(J/2)^2 + E^2} \Bigr)} 
                   \Biggr \} = \ln 3,
\label{tcdiced}
\end{eqnarray}
where $\beta_c = 1/(k_{\rm B} T_c)$ and $T_c$ denotes the critical temperature.

\section{Numerical results and discussion}

Before proceeding to a discussion of the most interesting numerical results it is worthwhile to mention 
that the results obtained in the preceding section hold for the ferromagnetic ($J > 0$) as well as antiferromagnetic ($J < 0$) nearest-neighbour interaction $J$. For simplicity, we shall restrict ourselves 
in what follows to the latter case corresponding to the ferrimagnetically ordered system. It is noteworthy, however, that the critical frontiers of the ferromagnetic model are congruent with those displayed 
below for the ferrimagnetic model as a result of an invariance of the critical condition (\ref{tcdiced}) 
with respect to the transformation $J \to -J$. It should be also mentioned that the critical condition 
(\ref{tcdiced}) does not depend on a sign of the parameter $E$, hence, one also may consider without 
loss of generality $E \geq 0$.

\subsection{Ground-state phase diagrams}

Let us firstly focus on the ground-state phase diagrams of the system under investigation. As the 
easy-plane uniaxial single-ion anisotropy ($D < 0$) reinforces, a first-order phase transition from the ferrimagnetically ordered phase towards the disordered phase should be expected to occur owing to energetic favoring of a non-magnetic state $S = 0$ of the spin-1 atoms. It is quite interesting to ascertain, however, that the biaxial single-ion anisotropy stabilizes the ferrimagnetically ordered ground state in that it enhances a critical value of the uniaxial single-ion anisotropy at which both ferrimagnetic as well as disordered phases coexist together:
\begin{eqnarray}
\frac{D_c}{|J|} = \frac{-3}{2 \sqrt{1 - (E/D)^2}}.
\label{cc}
\end{eqnarray} 
This remarkable feature is in an obvious contradiction with naive expectations, since the biaxial term 
$E$ entails uprise of local quantum fluctuations that manifest themselves in a quantum reduction of the sub-lattice magnetization $m_B$. Actually, arbitrary but non-zero biaxial single-ion anisotropy destroys a perfect ferrimagnetic order even at zero temperature and the occurring quantum ferrimagnetic 
phase can be characterized through its sub-lattice order parameters given by:
\begin{eqnarray}
m_A = - \frac{1}{2}, \qquad \qquad m_B = \frac{3}{\sqrt{9 + 4 (E/|J|)^2}}
\label{MGS}
\end{eqnarray} 

As far as the influence of next-nearest-neighbour interaction $J'$ is concerned, one should vigorously distinguish between the effect of ferromagnetic ($J' > 0$) and antiferromagnetic ($J' < 0$) interaction. 
The former interaction aligns ferromagnetically the next-nearest-neighbouring spin-1/2 atoms and thus, 
the spin system remains long-range-ordered independently of a strength of the uniaxial single-ion anisotropy 
that acts purely on the spin-1 sites. On the other hand, the latter interaction causes a spin frustration between the next-nearest-neighbours and in the consequence of that, it acts in accompliance with the uniaxial single-ion anisotropy in view of destroying the ferrimagnetic long-range order. This synergetic effect 
can also be viewed from a modified version of the critical condition (\ref{cc}) when switching on the next-nearest-neighbour interaction $J'$: 
\begin{eqnarray}
\frac{D_c}{|J|} = \frac{-3}{2 \sqrt{1 - (E/D)^2}} - \frac{J'}{2|J|} 
\quad (\mbox{for} \hspace{2mm}  J' < 0 \hspace{2mm} \mbox{only}).
\label{ccd}
\end{eqnarray}

\subsection{Finite-temperature phase diagrams}

Now, let us turn to a detailed discussion dealing with the finite temperature phase diagrams. For this purpose, figure 2 shows the variation of reduced critical temperature with the uniaxial single-ion anisotropy 
for several fixed values of the ratio $E/D$ and $J'/|J| = 0.0$. The quantum ferrimagnetic phase can be located inside of the depicted phase boundaries, while above them the usual paramagnetic phase becomes 
\begin{wrapfigure}{i}{0.6\textwidth}
\vspace{-8mm}
\centerline{\includegraphics[width=0.58\textwidth]{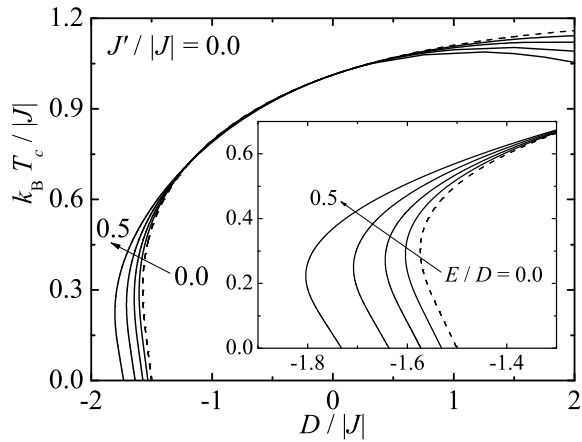}}
\vspace{-7mm}
\caption{The reduced critical temperature against the uniaxial single-ion anisotropy at 
$E/D = 0.0$, 0.2, 0.3, 0.4, 0.5 in ascending order along the direction of arrows. 
The case $E/D = 0.0$ is plotted as a dashed line for visual effect.}
\vspace{-5mm}
\label{fig2}
\end{wrapfigure}
stable. There are two interesting things to observe here. Firstly, there is a slight 
suppression of the critical temperature in the region of easy-axis uniaxial  
single-ion anisotro\-pies ($D > 0$) for any $E \neq 0$. 
This unusual feature can be interpreted in terms of strengthening of the quantum fluctuations caused by the non-zero biaxial single-ion anisotropy. Actually, the greater the ratio $E/D$, the more impressive reduction of $T_c$ can be observed. Secondly, the insert of figure 2 clearly demonstrates that the system exhibits interesting reentrance in the vicinity of the critical value $D_c/|J|$ given by the condition (\ref{cc}).
For the uniaxial anisotropies slightly below this critical value, one starts from the disordered ground 
state before entering the ordered phase at lower critical temperature and finally, the ferrimagnetic long-range order disappears due to thermal fluctuations at the upper critical temperature. Notice that such pure reentrant transition with two consecutive critical points have not been reported in the analogous mixed-spin system on the honeycomb lattice \cite{Str04}. This observation would suggest that a presence 
of the reentrant transition can possibly be attributed to higher coordination number of the spin-1/2 sites.  

The influence of the next-nearest-neighbour interaction on the phase diagrams is illustrated in figure 3. 
As one can see, the spin system remains long-range ordered for arbitrary strength of the ferromagnetic 
next-nearest-neighbour interaction ($J' > 0$).  
\begin{wrapfigure}{i}{0.6\textwidth}
\vspace{-6mm}
\centerline{\includegraphics[width=0.58\textwidth]{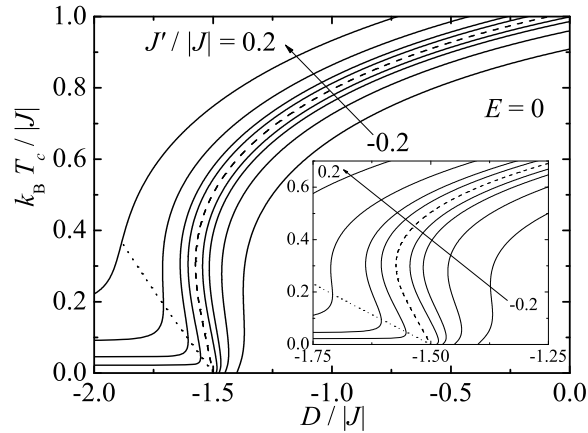}}
\vspace{-7mm}
\caption{The reduced critical temperature against the uniaxial single-ion anisotropy at 
$J'/|J| = -0.2$, -0.1, -0.05, -0.025, 0.0, 0.025, 0.05, 0.1, 0.2 in ascending order along 
the direction of arrows. The case $J'/|J| = 0.0$ is plotted as a dashed line for visual effect.
A dotted line shows the line of compensation temperatures.}
\vspace{-2mm}
\label{fig3}
\end{wrapfigure}
When small ferromagnetic coupling $J'$ is taken into consideration, 
however, the critical temperature versus uniaxial anisotropy dependence can be characterized by the 
$S$-shaped curve indicating an occurrence of the reentrant transitions with three consecutive critical 
points (see the insert in figure 3). It is noteworthy that this reentrant behaviour gradually vanishes 
as the ratio $J'/|J|$ strengthens. Even more interesting situation emerges when assuming the
antiferromagnetic next-nearest-neighbour interaction ($J' < 0$). In the latter case, the displayed 
critical lines imply an existence of reentrant transitions with either two or three consecutive 
critical points before approaching $D_c/|J|$ given by equation (\ref{ccd}). Notice that a range of 
the critical boundaries corresponding to the reentrant transition with two (three) critical temperatures 
overwhelms at relatively smaller (greater) values of $|J'|/|J|$, while the reentrance completely disappears 
from the critical lines above certain boundary value of $|J'|/|J|$. Finally, it is worth mentioning that 
similar influence of the next-nearest-neighbour interaction on the criticality is observable even 
if a non-zero strength of the biaxial single-ion anisotropy would be considered.

\subsection{Temperature dependence of magnetization}

In this part, the temperature dependence of magnetization (to be further abbreviated as $m(T)$ curves) 
will be explored in particular. When both kinds of atoms are stochiometrically taken into account, 
the resultant magnetization reduced per one lattice site of the mixed-spin diced lattice can be 
defined as $m = |m_A + 2m_B|/3$. Figure 4(a) depicts the effect of uniaxial single-ion anisotropy 
on the shape of $m(T)$ curves. 
\begin{figure}
\vspace{-9mm}
\centerline{\includegraphics[width=0.9\textwidth]{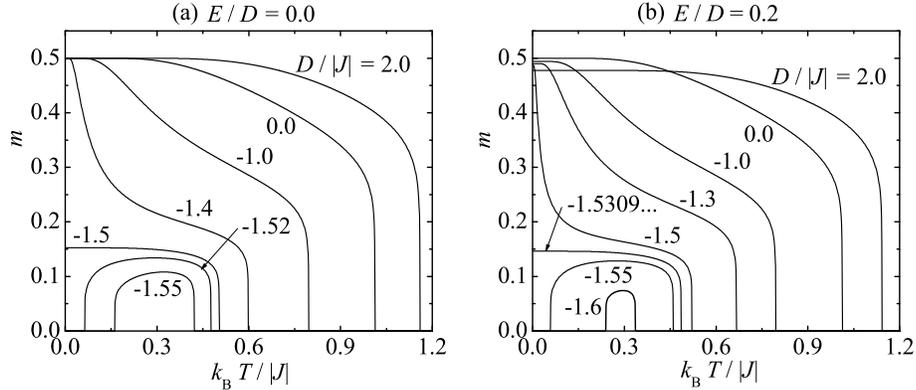}}
\vspace{-9mm}
\caption{Temperature variation of the resultant magnetization as the uniaxial single-ion anisotropy varies
and the ratio $J'/|J| = 0.0$. The parameter of biaxial single-ion anisotropy is fixed to be: 
(a) $E/D = 0.0$; (b) $E/D = 0.2$.}
\label{fig4}
\end{figure}
The downward curvature observed for strong negative uniaxial anisotropies can obviously be attributed 
to a more rapid thermal variation of the sub-lattice magnetization $m_B$, which is thermally more 
easily disturbed than $m_A$ under the condition $D < 0$. In accordance with the phase diagrams 
displayed earlier, one finds $m(T)$ curves exhibiting the reentrant transitions with two consecutive 
critical points slightly below the critical value $D_c/|J|$ given by the condition (\ref{cc}). 
For comparison, figure 4(b) illustrates how the biaxial single-ion anisotropy affects a temperature 
variation of the resultant magnetization when the ratio $E/D = 0.2$ is fixed and $D/|J|$ varies. 
It is quite evident that the most ultimate difference between the $m(T)$ curves depicted in 
figures 4(a) and 4(b), respectively, rests in the zero-temperature limit of the latter ones 
that do not start from their saturation value owing to the quantum reduction of the sub-lattice 
magnetization $m_B$ caused by the biaxial anisotropy.

Next, the effect of next-nearest-neighbour interaction $J'$ upon the shape of $m(T)$ curves 
will be clarified. The effect of biaxial anisotropy is neglected here for simplicity, however, 
it worthy to note that the curves displayed below remain qualitatively unchanged even if the 
biaxial anisotropy term would be non-zero. Figure 5(a) shows typical $m(T)$ curves under 
the assumption of small ferromagnetic interaction 
\begin{figure}
\vspace{-9mm}
\centerline{\includegraphics[width=0.9\textwidth]{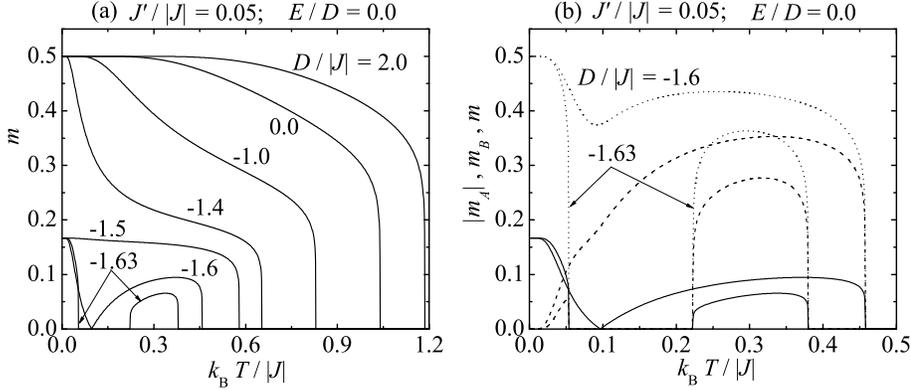}}
\vspace{-9mm}
\caption{The effect of small ferromagnetic next-nearest-neighbour interaction $J'/|J| = 0.05$ on 
the temperature dependence of magnetization. Figure 5(a) shows the variation of the resultant magnetization, 
Figure 5(b) depicts in addition to the total magnetization (solid lines) also the temperature variation 
of the sub-lattice magnetization $m_A$ (dotted lines) and $m_B$ (dashed lines), respectively.}
\label{fig5}
\end{figure}
$J'/|J| = 0.05$. The most remarkable differences compared to the $m(T)$ curves discussed 
earlier can be found in the vicinity of the boundary value $D_c/|J| = -1.5$, which is strong 
enough to force all the spin-1 atoms towards their non-magnetic state $S = 0$ in the 
ground state. For illustration, figure 5(b) depicts the sub-lattice magnetization of two most
striking dependences in addition to the resultant magnetization. As one can see, there 
appears $m(T)$ curve with one compensation point if the uniaxial anisotropy is chosen 
slightly below its boundary value (for instance $D/|J| = -1.6$).
The compensation phenomenon arises on account of thermally induced increase of the sub-lattice 
magnetization $m_B$, which overwhelms the sub-lattice magnetization $m_A$ above the 
compensation temperature. When the uniaxial anisotropy is taken to be even more negative, there emerges
$m(T)$ curve with three consecutive critical points instead of the one with the compensation point
(the case $D/|J| = -1.63$). Notice that further decrease of the uniaxial single-ion anisotropy shrinks 
the temperature range between the second and third critical point until the ordered phase completely 
vanishes from this higher-temperature region. 

At last, we shall briefly discuss the effect of small antiferromagnetic next-nearest-neighbour interaction 
($J'/|J| = -0.05$) on the temperature dependence of the magnetization as shown in figure 6. According to
\begin{figure}
\vspace{-9mm}
\centerline{\includegraphics[width=0.9\textwidth]{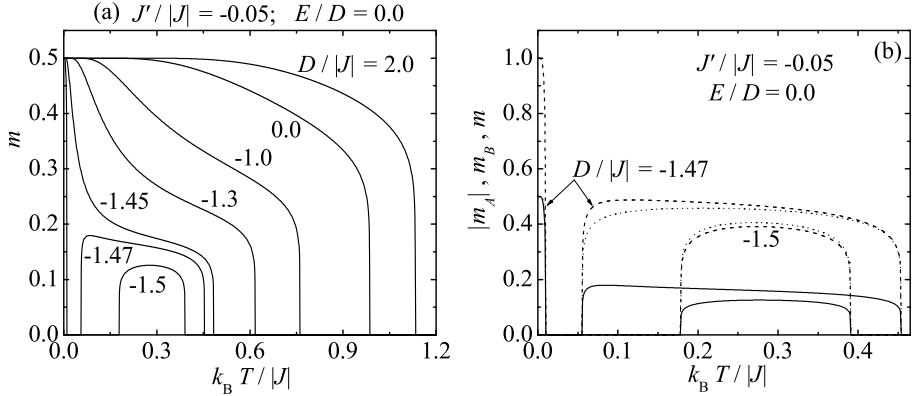}}
\vspace{-9mm}
\caption{The same as in Figure 5 but for the case of small antiferromagnetic next-nearest-neighbour 
interaction $J'/|J| = -0.05$.}
\label{fig6}
\end{figure}
equation (\ref{ccd}), the critical line terminates due to the spin frustration already at $D_c/|J| = -1.475$.
It is quite apparent from figures 6(a) and 6(b) that the magnetization exhibits two reentrant transitions 
for uniaxial anisotropies slightly below this critical value, for example at $D/|J| = -1.5$.
Contrary to this, the resultant magnetization may show the reentrant transition with three critical temperatures when the uniaxial anisotropy is selected slightly above this boundary value 
(the case $D/|J| = -1.47$). 
Whether the reentrant transition with two or three critical points is observed, however, depends very sensitively on the relative magnitude of all interaction parameters. It appears worthwhile to remark
that both kinds of reentrant transitions disappear as the magnitude of next-nearest-neighbour interaction 
is considered to be strong enough.

\section{Concluding Remarks}

In the present article, the mixed spin-1/2 and spin-1 Ising model on the diced lattice has exactly been 
treated within the generalized star-triangle mapping transformation. The main focus of the present work 
has been aimed at the influence of the next-nearest-neighbour interaction, uniaxial and biaxial 
single-ion anisotropies on the magnetic behaviour of the system under investigation. It has been 
found that the single-ion anisotropy may possibly cause the reentrant transition with two critical 
temperatures only. On the other hand, the ferromagnetic next-nearest-neighbour interaction may lead just 
to the reentrant transition with three consecutive critical points, whereas the antiferromagnetic next-nearest-neighbour coupling may be considered as possible cause of both kinds of reentrant transitions. 

Another interesting finding to emerge here consists in the fact that the biaxial single-ion anisotropy 
has been confirmed to be a possible source of the quantum reduction of magnetization. Our further
observations suggest that the biaxial anisotropy cannot qualitatively change the character of reentrant transition caused by the uniaxial single-ion anisotropy term or the next-nearest-neighbour coupling. 
Finally, we should remark that a combined effect of uniaxial and biaxial single-ion anisotropies 
with the next-nearest-neighbour interaction is responsible for manifold temperature dependences 
of the resultant magnetization by assuming the ferrimagnetically ordered spin system.

%
%
\label{last@page}
\end{document}